\documentclass[12pt,preprint]{aastex} 
\usepackage{graphicx} 
\usepackage{natbib}

\newcommand{\skipthis}[1]{}

\def\nh3{$\rm{NH_3}$}
\def\NH3{$\rm{NH_3}$} 
 
\def\msun{M$_\odot$}
 
\def\lsun{L$_\odot$} 
\def\kms-1{km~s$^{-1}$}
\def\h2o{$\rm{H_2O}$} 
\def\h2{$\rm{H_2}$} 
\def\CM2{$\rm{cm^{-2}}$}
\def\cm3{$\rm{cm^{-3}}$}

\def\n2h+{N$_2$H$^+$}
\def\ch3oh{CH$_3$OH}

\newcommand{\lsim}{${\raisebox{-.9ex}{$\stackrel{\textstyle<}{\sim}$}}$ }

\begin{document}

\title{\ Fragmentation at the Earliest Phase of Massive Star Formation}

\author{Qizhou Zhang, Yang Wang, Thushara Pillai, Jill Rathborne}

\affil{Harvard-Smithsonian Center for Astrophysics, 60 Garden Street, Cambridge, MA 02138}

\email{qzhang@cfa.harvard.edu}


\begin{abstract}
We present 1.3mm continuum and spectral line images of two massive molecular clumps
P1 and P2 in the G28.34+0.06 region with the Submillimeter Array. 
While the two clumps contain masses of 1000 and 880 \msun, 
respectively, P1 has a luminosity $< 10^2$ \lsun, and a lower gas 
temperature and smaller line width than P2. Thus, P1 appears to be at a much
earlier stage of massive star formation than P2. The high resolution SMA observations reveal
two distinctive cores in P2 with masses of 97 and 49 \msun, respectively. 
The 4 GHz spectral bandpass  captures line emission
from CO isotopologues, SO, CH$_3$OH, and CH$_3$CN, similar to hot
molecular cores harboring massive young stars. The P1 clump, on the
other hand, is resolved into five cores along
the filament with masses from 22 to 64 \msun\ and an average
projected separation of 0.19 pc.  Except $^{12}$CO, no molecular line
emission is detected toward the P1 cores at 
a 1$\sigma$ rms of 0.1 K. Since strong $^{12}$CO and C$^{18}$O emissions are seen
with the single dish telescope at a resolution of 11$''$, the non-detection
of these lines with the SMA indicates a depletion factor upto $10^3$.
While the spatial resolution of the SMA is better than the expected Jeans length, the 
masses in P1 cores are much larger than the thermal Jeans mass, indicating
the importance of turbulence and/or magnetic fields in cloud fragmentation.
The hierarchical structures in the P1 region provide a glimpse of the initial
phase of massive star and cluster formation.

\end{abstract}

\section{Introduction}

In the past decade, significant progress has been made toward
understanding the more evolved stages of massive star formation, 
phases from hot molecular cores to hyper compact HII regions. 
It becomes clear that massive
proto B stars are associated with molecular 
outflows and rotating disks \citep{cesaroni1997,zhang1998b,zhang2001,beuther2002b}.
Therefore, they likely
form through a disk mediated accretion \citep{zhang2005b,cesaroni2007}. 
Active mass accretion can continue during the hyper compact HII region phase when the 
ionization radius is within the gravitational radius \citep{keto2002a,keto2002b,sollins2005a}. 
All these studies represents
phases of massive star formation after the central star has formed and 
begun hydrogen burning. Contrary to these progresses,
little is known about the initial phases of massive star formation. 
Since most massive stars are born in a cluster environment \citep{lada2003},
the initial phase of
massive star formation is intimately linked to cluster formation and
the fragmentation of molecular clouds.
While theoretical work provides physical insights into fragmentation
process, direct observations of precluster molecular clouds
are crucial in revealing how the process may take place.
Among all the observational efforts, studies of infrared dark clouds (IRDC), 
massive clouds at extremely early stages of evolution, hold great promise
and has grown rapidly over the past several years.

IRDCs were first discovered in large number by the Infrared Space
Observatory (ISO) and the Midcourse Space Experiment (MSX)
\citep{egan1998, carey1998, carey2000, hennebelle2001, simon2006a, simon2006b}
through infrared absorption against the bright galactic background.
Systematic
studies found that many IRDCs contain over $10^3$ \msun\ of 
dense molecular gas \citep{simon2006b, rathborne2006, pillai2006b}. While
some massive IRDCs show signs of massive star formation
through \h2O maser emission \citep{wang2006}, CH$_3$OH maser 
emission \citep{ellingsen2006},
molecular outflows \citep{beuther2007a},
and bright 24 $\mu$m emission \citep{rathborne2005}, a
large majority do not show signs of star formation \citep{wang2006}. 
With a typical star formation
efficiency, the large amount of dense gas in these IRDCs
makes them the natural birth place for massive stars and clusters.
Thus, these regions are premium sites for uncovering
the initial conditions of massive star and cluster formation.

\citet{wang2008} imaged an IRDC G28.34+0.06 with the
Very Large Array in the \nh3 (1,1) and (2,2) lines in the D configuration.
An overview of the region is presented in Figure 1. The cloud, 
at a distance of $\sim$4.8 kpc, contains several $10^3$ M$_\odot$ of
dense gas along the infrared absorption filament extending 6 pc
in the sky \citep{carey2000,
rathborne2006, pillai2006b}. Two prominent dust continuum clumps, P1 and P2,
are revealed  in the 850 and 450 $\mu$m images obtained from the JCMT
\citep{carey2000} and the 1.2mm image obtained from the
IRAM 30m telescope \citep{rathborne2006}.
Despite the fact that two clumps contain a similar amount of dense gas 
within 0.3 pc, the P2 region has a high
gas temperature of 30 K, large \nh3 line width of 3.3 \kms-1\,
and is associated with
a strong 24 $\mu$m point source with far IR luminosity of $10^3$ \lsun.
This is in contrast to P1 which has a gas temperature of 13 K, a relatively
narrow \nh3 line width of 1.7 \kms-1, and an upper limit to the luminosity of
$10^2$ \lsun \citep{wang2008}. Furthermore, the gas in P1 appears to be 
externally heated with
temperatures decreasing from 20 K in the outside of the cloud
to 13 K inside of the cloud. Likewise, the turbulence measured by the
\nh3 line widths
appears to decrease from larger spatial scales to smaller scales. 
These observations led \citet{wang2008} to suggest that P1 is at a much 
earlier stage of massive star formation compared to P2.

In order to study the cloud structure,
we imaged the P1 and P2 clumps at $ \sim 1''$ resolution with the SMA. 
The sensitive continuum image at 1.3mm reveals 5 dense cores toward
the P1 clump.  
The P2 region exhibits rich molecular line emission similar
to hot molecular cores, while the P1 cores are absent of molecular line
emission, which indicates heavy depletion.

\section{Observations}

Observations with the SMA\footnote{The Submillimeter 
Array is a joint project between the Smithsonian
Astrophysical Observatory and the Academia Sinica Institute of Astronomy and
Astrophysics, and is funded by the Smithsonian Institution and the Academia
Sinica.}  were carried out with 7 antennas
in the compact configuration on 2006 August 07, and 8 antennas
in the extended configuration on 2008 July 22. The pointing centers 
of the observations were RA (J2000) = 18:42:52.09, Dec (J2000) = 
-3:59:52.00 for P2
and RA (J2000) = 18:42:50.74, Dec (J2000) = -04:03:15.34 for P1.
The FWHM of the primary beam of the two pointings is shown as
the dashed circles in Figure 1.

For the compact configuration, quasars 1741-038 and 1908-201 
were used as time dependent gain calibrators, Uranus and
Ganymede were used as bandpass and flux calibrators, respectively. 
The system temperatures during the observations varied
from 200 to 600 K. 
The receivers were tuned to an LO frequency of 225 GHz. With IF frequencies 
of $4 - 6$ GHz, the observations covered rest frequencies from
220 through 222 GHz in the LSB, and
230 through 232 GHz in the USB, with a uniform channel spacing of 
0.8125 MHz ($\sim$ 1 \kms-1) across the entire band. 

For observations with the extended configuration, we used 1751+096 and 1911-201
as gain calibrators, 3C279 and 3C454.3 as bandpass calibrators, and Callisto
as a flux calibrator. The system temperatures during the observations 
were around 100 K. The receivers were tuned to an LO frequency of 220 GHz
and covered rest frequencies of
215.7 through 217.7 GHz in the LSB, and
225.7 through 227.7 GHz in the USB, with a uniform channel spacing of 
0.4062 MHz ($\sim$ 0.5 \kms-1) across the entire band.
A detailed description of the SMA is given in \citet{ho_moran_lo2004}.

The visibility
data were calibrated with the IDL superset MIR package developed for the
Owens Valley Interferometer.  The absolute flux level is accurate to about 15\%.
After the
calibration in MIR, the visibility data were exported to the MIRIAD format for
further processing and imaging.  The continuum is constructed from line
free channels in the visibility domain. The projected baselines range 
from 12 to 68 m in the compact configuration, and 20 to 221 m in the
extended configuration.
We combined the continuum data from both configurations, which yields
a synthesized beam of about 1.2$''$, and a $1\sigma$ rms of 1 mJy
in the naturally weighted maps.
The spectral line data presented here is from the compact configuration
only, with a $1\sigma$ rms of 90 mJy~beam$^{-1}$ per 1.2 \kms-1 channel, 
and a spatial resolution of $3''$.  The uncertainty in the 
absolute position
is $\lsim 0''.2$, derived from comparing the secondary gain calibrators with
their catalog positions.

\section{Results and Discussions}

\subsection{Continuum Emission}

Figure 2 presents the 1.2mm continuum emission from the IRAM 30m 
telescope \citep{rathborne2006}, 
the \nh3 (1,1) emission from the VLA \citep{wang2008}, and the SMA 1.3mm dust 
continuum emission. To describe the dust emission, we follow the nomenclature in the 
literature that {\it clumps} refer to structures with sizes of $\sim$ 1 pc, {\it cores}
refer to structures within a clump with sizes of $\sim$ 0.1 pc,  and {\it condensations}
refer to sub structures within a core. As shown in Figure 2, 
the two dust clumps P1 and P2
detected with the IRAM 30m telescope and JCMT at resolutions 
from 11$''$ to 15$''$ \citep{rathborne2006,carey1998} are resolved
by the SMA into
multiple structures at $1''.2$ resolution. The P1 clump splits
into five cores (P1-SMA1, P1-SMA2, P1-SMA3, P1-SMA4 and P1-SMA5)
with an average projected separation of $8''$ or 0.19 pc. These cores
show further substructures that may correspond to condensations.
Since we do not have complementary spectral line data at a resolution of $1''$,
we defer analysis of the substructure to a future paper.
The emission in P2, on the other hand, is resolved into two strong,
distinctive cores, P2-SMA1 and P2-SMA2. \h2O
masers reported in  \citet{wang2006} coincide
with P1-SMA2 and  P2-SMA2 to within 0.1$''$. 

The continuum images from the SMA and the IRAM 30m telescope
are in a remarkably good agreement with the \nh3 emission. The five
continuum cores (P1-SMA1 through P1-SMA5) lie along the \nh3 ridge.
There is a feature in the 30-m continuum image extending toward 
another \nh3 peak in the northwest. The SMA image from the compact configuration
reveals a peak at 5 mJy ($4 \sigma$). This peak
is not detected in the $1''.2$ resolution image presented in Figure 2.

The 1.2mm integrated flux densities from the IRAM 30m telescope are 2.6 Jy and
1.3 Jy, toward P2 and P1, respectively. Neither clumps have
detectable continuum emission at centimeter wavelength at a level 
of 1 mJy \citep{wang2008},
thus the mm flux arises predominantly from the dust emission.
Using the dust opacity 
of \citet{hildebrand1983}, an emissivity index of $\beta$ = 1.5, 
and a temperature of 13 and 30 K derived from \nh3 \citep{wang2008}, we 
obtained  dust mass of 1000 and 880 \msun\ for P1 and P2, respectively, 
with an average \h2 density of $3 \times 10^5$ \cm3. 
The masses in the P1 and P2 clumps are consistent with the estimate
in \citet{carey2000} from 850 $\mu$m and 450 $\mu$m if one factors
in the different values of dust temperatures used.
For $\beta$ = 2 as assumed in \citet{rathborne2006}, we obtain
a mass of 2000 and 1760 \msun\ for the P1 and P2 clumps, respectively.

We measure the flux density toward the dust cores in the SMA images.
For the P2 cores, the integrated fluxes were obtained from fits
of elliptical Gaussian. For the P1 cores, we derived the integrated fluxes
within a polygon encompassing each core.
The peak positions and the integrated fluxes are
listed in Table 1. 
The \nh3 (1,1) and (2,2) data reveal
gas temperatures of 13 to 16 K in cores toward P1, and 30 K in cores 
toward P2. Using the same dust opacity law and a dust
emissivity index of $\beta$ = 1.5, appropriate for
massive dense cores based on multi-wavelength observations at mm and submm
wavelengths \citep{beuther2007b}, we obtained dust masses for the cores
ranging from 22 to 97
\msun. The corresponding \h2 densities in these cores range from $10^6$ to 
$10^7$ \cm3. These parameters are also listed in Table 1.

\subsection{Line Spectra and Depletion}

Figure 3 shows the spectra over the 4 GHz bandpass of the SMA toward 
P2-SMA1, and the strongest core, P1-SMA4, in the P1 clump. The spectra are made
from the compact configuration with a resolution of $3'' - 4''$.
Toward P2-SMA1, 
emissions of CO isotopologues, SO, and more complex molecules
CH$_3$OH, H$_2^{13}$CO, HNCO,
as well as CH$_3$CN are detected. The presence of complex organic
molecule is similar to that of hot cores observed
with the SMA \citep{beuther2005a,zhang2007a,rathborne2008}.
The classical hot core molecule CH$_3$CN is detected in the K=0, 1, 2, 3, 
and 4 components of the J=12-11 transition.
Physical parameters, namely, column density and
temperature, have been estimated by fitting CH$_3$CN spectrum, 
in the LTE approximation,
and taking into account the optical depth effect. 
The best fit gives a temperature of 120 K, representative of the temperature
value in a typical hot core.

In contrast to the rich line spectra in P2-SMA1, no molecular line 
emission except the faint $^{12}$CO 2-1 emission is detected 
toward P1-SMA4. One can estimate CO depletion from the 2-1
emission of the CO isotopologes. Toward the P2 position,
the $^{13}$CO to C$^{18}$O ratio yields an optical depth of 1.8 in the
$^{13}$CO 2-1 line assuming a [$^{13}$CO/$^{18}$O] of 40.
We use the C$^{18}$O 2-1 transition to estimate the
CO column density because it is optically thin, and is least affected
by missing short spacing in the SMA data as compared to $^{13}$CO
and $^{12}$CO. With a brightness temperature of 2.5K, a
FWHM of 2.5 \kms-1 in the C$^{18}$O 2-1 line width, and
a gas temperature of 30K, we obtain a $^{12}$CO column density of 
$4 \times 10^{14}$ \CM2, assuming a [$^{12}$CO/C$^{18}$O]
of 240. On the other hand, the \h2 column density
derived from the dust emission, assuming a dust to gas ratio of 1:100
amounts to $8 \times 10^{23}$ \CM2. These values give rise to
a [CO/H$_2$] ratio of $1.2 \times 10^{-6}$, a factor of 100 lower
than the standard value of $10^{-4}$ in [CO/H$_2$] ratio.

Similarly, one can estimate CO depletion from the upper
limit of 0.1K in the C$^{18}$O 2-1 line in P1-SMA4.
If the C$^{18}$O 2-1 emission is optically thin, the 0.1 K
upper limit and the 13K gas temperature yield a $^{12}$CO
column density of $2.2 \times 10^{16}$ cm$^{-2}$. The 1.3 continuum 
emission gives rise to an \h2 column density of $2.1 \times 10^{23}$ cm$^{-2}$.
Therefore, we obtain a [CO/H$_2$] ratio of $\sim 10^{-7}$,
or a CO depletion of $10^3$ from the standard [CO/H$_2$] ratio.

The non detection of molecular line emission with the SMA is in 
staunch contrast with
the strong line emission in CO isopologues, CS, HCN toward P1
with the IRAM 30m telescope (Rathborne personal communication). 
The brightness temperatures 
of the $^{13}$CO and C$^{18}$O 2-1 lines are 3 and 1 K, respectively. 
The line ratio yields
an optical depth in $^{13}$CO of 2.  With a gas temperature of 16 K
from the single dish \nh3 \citep{pillai2006b} and a brightness temperature of
1K in the C$^{18}$O 2-1 line, we obtain a column density of
$7.5 \times 10^{17}$ cm$^{-2}$ in $^{12}$CO. The \h2\ column density, 
based on the measurement from the IRAM 30m 
telescope at 1.2mm is $3 \times 10^{23}$ cm$^{-2}$. 
These values yield a [CO/H$_2$] ratio of $2.5 \times 10^{-6}$,
which is 40 times lower than the typical value (10$^{-4}$)
in the Galaxy.  \citet{pillai2007} reported a similar results in the 
[CO/H$_2$] ratio toward this region.

The estimates above yield  CO depletion upto 10$^2$ to $10^3$ in
P2 and P1, respectively.
It is known that carbon bearing molecules such as
CO are heavily depleted in low-mass prestellar cores \citep{bergin2007}.
This phenomenon is seen in massive cores over a scale of nearly 0.1 pc,
much larger than the $10^3$ AU scale in the low-mass prestellar cores.
The timescale of depletion for an average density of $10^5$ \cm3\
is several $10^4$ yrs \citep{bergin2007}. This timescale sets a 
lower limit to the age of the P1 cores at $> 10^5$ years.

\subsection {Core Structure and Dynamical State}

The dust emission from the SMA observations appears to be centrally
peaked. For a core internally
heated by a protostar, the dust temperature scales as
$T_{dust} \propto r^{-a}$
\citep{scoville1976} with $a = 0.33$. If the core density $\rho$ 
scales as $r^{-b}$, then
the flux density from dust emission $F \propto \int \rho T_{dust} ds$,
where $s$ is the length along the line of sight. When $a + b > 1$,
we find $F \propto r^{-(a + b - 1)}$. 
The Fourier transform of the flux density becomes
$ A \propto S_{uv}^{(a + b - 3)}$ \citep{looney2000, beuther2007b},
where $A$ is the visibility amplitude
and $S_{uv} = \sqrt{(u^2 + v^2)}$ is the UV distance.

Figure 4 presents the density profile for P2-SMA1, and P1-SMA4, two
of the strongest cores in the clumps.
The visibility is vector averaged over concentric annuli around
the center of the core. The least-squared fit to visibilities
yields $b = 1.6 \pm 0.2$ for P2-SMA1, and $b = 2.1 \pm 0.2$ for P1-SMA4,
if $a = 0.33$. The visibility data toward the P1-SMA4 have larger scatters
due to a limited signal-to-noise ratio, thus the uncertainty to the power-law
index is larger than the statistical error quoted here.

The density profiles in both P1 and P2 cores are consistent
with a power law relation $r^{-b}$ with $b$ between 1.5 to 2, similar
to those found in hot molecular cores \citep{vandertak2000a}.
Such power law structures suggest that these cores may be evolved from
an equilibrium state. 
A self gravitating core in a hydrostatic equilibrium has
a density structure of $r^{-2}$. A hydrostatic core truncated by
external pressure, $i.e.$, the Bonnor-Ebert sphere has a similar density
structure in the outer radii with a flatter inner profile. These density
structures have been found in low-mass prestellar cores \citep{alves2001}.
The fact that the more massive cores in the G28 region shows similar
structures is rather intriguing, and suggests that
cores may evolve slowly to allow them reaching a quasistatic equilibrium state.
This is consistent with the age of the cores
inferred from the time scale of chemical depletion.

\subsection{Core Fragmentation and Cluster Formation}

The early evolutionary stage of the P1 clump makes it an excellent target 
to study the initial fragmentation in molecular clouds. The structures
revealed by the high resolution SMA observations are consistent with a hierarchical
fragmentation of molecular clouds. It appears that the initial fragmentation 
gives rise to
5 massive cores with masses from 22 to 64 \msun\ with an average separation
of $8''$ or 0.19 pc. The cores appear to further fragment 
as core densities continue to increase.
The average density in the P1 clump based on the IRAM 30m telescope is $3 \times
10^5$ \cm3. With an average gas temperature in the clump of 16 K, we find a Jeans mass
of 1 \msun\ and a Jeans length of 0.05 pc. Despite the fact that the SMA
observations spatially resolve the Jeans length, the core
masses are more than a factor
of 10 larger than the thermal Jeans mass. The large discrepancy
between the Jeans mass and the observed core mass
indicates that cloud fragmentation may not be controlled solely by the
thermal pressure
and gravity. Other stabilizing factors are required to account for the large mass
observed. 
\citet{wang2008} reported a systematic decrease in the \nh3 line width
from large scales to small scales of $3'' - 4''$, which cannot be attributed to
organized motion. This decrease in line widths is interpreted as a 
turbulence decay in dense molecular gas. The observed \nh3 line width in the P1 cores
is 1.7 \kms-1, a factor of 8 larger than the thermal line width. This
supersonic turbulence likely plays an important role in fragmentation. 
The Viral masses measured from the \nh3 line widths 
are 103 and 24 \msun\ in P2 and P1 cores, respectively. The general
agreement between the Virial mass and gas mass lends support to the
notion that turbulence dominates fragmentation.
In addition to turbulence, magnetic fields may also play
a role in suppressing fragmentation. However, without direct
measurements of field strengths, it is difficult to
assess its importance relative to turbulence.

Several theoretical ideas postulate how massive star and cluster 
formation may begin.
Observations find that massive stars and clusters form in higher density
and more turbulent regions of molecular clouds as compared to their low mass
counterparts. \citet{bonnell2002} propose in the competitive accretion model
that clouds fragment initially into cores of Jeans mass
of $\sim$ 0.5 \msun. These cores subsequently form 
low-mass protostars that accrete the distributed gas from a 
reservoir of material in the molecular clump. Prototars located near
the center of the gravitational potential accrete at a
higher accretion rate because of a stronger gravitational pull,
thus, experience a faster mass growth. This competitive accretion
model reproduces the stellar
IMF observed \citep{bonnell2004}. 

Alternatively, \citet{mckee2002} propose a turbulent accretion 
model, in
which stars form via a monolithic collapse of a molecular cloud. The
heating from the embedded protostars increases the gas temperature,
and thus, the Jeans mass. Therefore, the core mass function is similar 
to the stellar IMF \citep{krumholz2005a, krumholz2007}. 

The observed core masses in P1 appear to be much larger than the core mass
assumed in the
competitive accretion model \citep{bonnell2005}. The
$3 \sigma$ mass sensitivity of 2.5 \msun\ in this study
does not detect the 0.5 \msun\ Jeans cores. However,
the SMA observations have a spatial resolution better than the Jean length
(0.05pc) predicted in the initial fragmentation. The fact that the core masses are
much larger than the Jeans mass points to an inconsistency with
the theoretical model. In addition, these cores do not represent 
transient objects because of the large masses, high average
densities $> 10^6$ \cm3, and heavy chemical depletion. On
the other hand, the mechanism of increasing Jeans mass through stellar heating proposed 
by  \citet{krumholz2005a,krumholz2006} does not appear to be sufficient.
To reach a Jeans mass of 30 \msun\ at a density of 
$10^5$ \cm3 requires a gas temperature of 100K. 
This temperature is a factor of 7 times larger than the
gas temperature derived from the \nh3 emission.
Therefore, the increase in temperature is insufficient
to stop fragmentation in the P1 region. The observations
indicate that turbulence plays a crucial role in shaping
fragmentation in clouds.
It has been suggested that turbulence
dissipates rapidly and may not be sufficient
to support cores at small scales. The \nh3 data from the VLA observations
\citep{wang2008} demonstrate that despite turbulence decay, 
the line widths measured at $3'' - 4''$ scales appear to be large
enough to support the cores in P1. 

The core mass in the P2 clump is approximately a factor of 2 larger than the core masses 
in the P1 clump. It is quite possible that the mass of cores and
protostars in the P1 region will
grow through infall and become similar to those in the P2 region. 
Assuming that
the cores in P2 is a few $10^5$ yrs, the infall rate required for 
the core mass growth 
is a few times $10^{-4}$ \msun~yr$^{-1}$.  Such a high infall rate is consistent with
expected infall rate $a^3 \over G$, where $a = 1.7$ \kms-1\
is the turbulent line width,
and G is the gravitational constant. The mass accretion to the star, however,
is likely to be much smaller as constrained by the low luminosity of
$< 10^2$ \lsun. Our observations
of the P1 and P2 clumps suggest massive star formation through
a low to intermediate mass phase as seen in P1 where masses of
cores and protostars continue to grow via infall and accretion.
More observations of similar regions are needed to test this scenario.

\section {Conclusion}

We carried out SMA observations of two massive clumps P1 and P2 which are at very different evolutionary stages. The main findings are:

(1) The P1 clump is resolved into five cores at a projected separation of 0.19 pc, 
and masses much larger than the Jeans mass. The large masses indicates that 
turbulence and/or magnetic fields play important role in fragmentation;

(2) The CO is depleted by a factor up to $10^3$ in the P1 cores at a scale 
of $>$ 0.07pc, much larger than the depletion area in low mass cores;

(3) These findings indicate that the initial conditions assumed in 
current theoretical models should be amended in light of these observations.

\begin{table}[h]
\caption{Physical Parameters in Cores}
\begin{center}
\begin{tabular}{rlllcrrrr} \hline \hline
Name & R.A.(J2000)   & Dec.(J2000)  & S & T  & Mass  & $\Delta$V & M$_J^a$  &  M$_V^b$ \\
     &  ($^h$ $^m$ $^s$)   &  ($^\circ$ $'$ $''$) & (mJy) & (K) &  (\msun)  & (\kms-1) & (\msun) & (\msun) \\ \hline
P2-SMA1  & 18:42:51.98  & -03:59:54.0 & 0.26  & 30 & 97  &  3.5  &  0.45  & 103 \\	
P2-SMA2  & 18:42:52.12  & -03:59:54.2 & 0.13  & 30 & 49 &  3.5  &  0.45  & 103 \\\hline	
P1-SMA1 & 18:42:51.18  & -04:03:06.9 & 0.034 & 13 & 38   &  1.7  &  0.16  &  24 \\	 
P1-SMA2 & 18:42:50.82  & -04:03:11.3 & 0.030 & 16 & 27   &  1.7  &  0.16  &  24 \\	 %
P1-SMA3 & 18:42:50.58  & -04:03:16.1 & 0.029 & 13 & 32   &  1.7  &  0.28  &  24 \\	%
P1-SMA4 & 18:42:50.30  & -04:03:20.3 & 0.057 & 13 & 64   &  1.7  &  0.19   &  24 \\ 	%
P1-SMA5 & 18:42:49.82  & -04:03:25.3 & 0.020 & 13 & 22  &  1.7  & 0.34   &  24 \\ \hline	%
\end{tabular}
\tablenotetext{a}{Jeans masses are computed using the average density and temperature in the clumps.}
\tablenotetext{b}{Viral masses are computed using a core radius of 0.02pc.}
\end{center}
\end{table}

\newpage

Fig. 1: The integrated intensity of the combined \nh3 (1,1)
emission in white solid contours overlaid on the Spitzer 8 $\mu$m
image in logarithmic color scales \citep{wang2008}. The \nh3 
image is contoured at
10\% of the peak (1 Jy~beam$^{-1}$ $\times$ km s$^{-1}$). The star
symbols mark the 24 $\mu$m emission peaks observed with
MIPS/\emph{Spitzer}. The cross symbols mark water maser
emission detected with the VLA. The thin dashed line indicates the
50\% of the sensitivity level of the 7 pointing mosaic in \nh3.
The \nh3 data have a resolution of $5'' \times 3''$, shown as the
shaded ellipse at the lower-left corner of the panel. The thick dashed
circles mark the SMA fields observed in this work.

Fig. 2: The dust continuum emission at 1.2mm from the IRAM 30m telescope,
at 1.3mm (225 GHz) from the SMA, and \nh3 (1,1)
integrated flux toward P1 and P2 clumps. The 1.2mm continuum emission from
IRAM is contoured at 10\% of the dust peak (1.3 Jy~beam$^{-1}$ in P2, 
and 0.26 Jy~beam$^{-1}$ in P1). The 1.3mm continuum emission is contoured in 
steps of 20 mJy~beam$^{-1}$ for P2 and 5 mJy for P1.
The \nh3 emission is contoured in steps of 50 mJy~beam$^{-1}$~\kms-1.
The shaded ellipses at the lower left corner of
each panel is the resolution of the observations.

Fig. 3: The spectra from the 2GHz pandpass in the LSB and USB
toward the dust cores P1-SMA4 and P2-SMA1.

Fig. 4: Visibility amplitude averaged over a concentric annulus versus UV distance in the P1-SMA4 and P2-SMA1 cores.
The solid line is a least-squared fit to the data.

\newpage

\begin{figure}[h]
\includegraphics{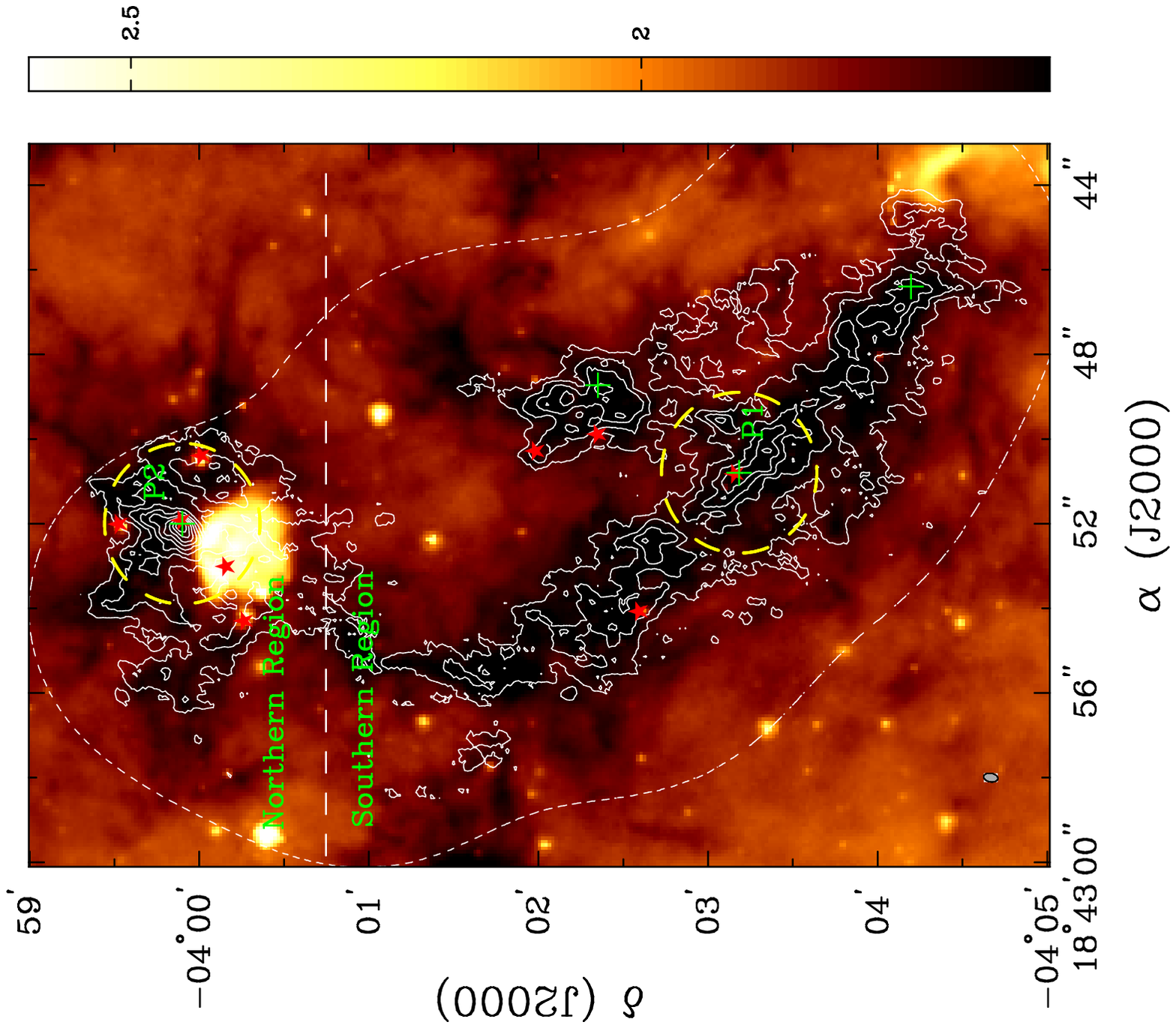}
\caption{ }
\end{figure}

\newpage

\begin{figure}[h]
\includegraphics{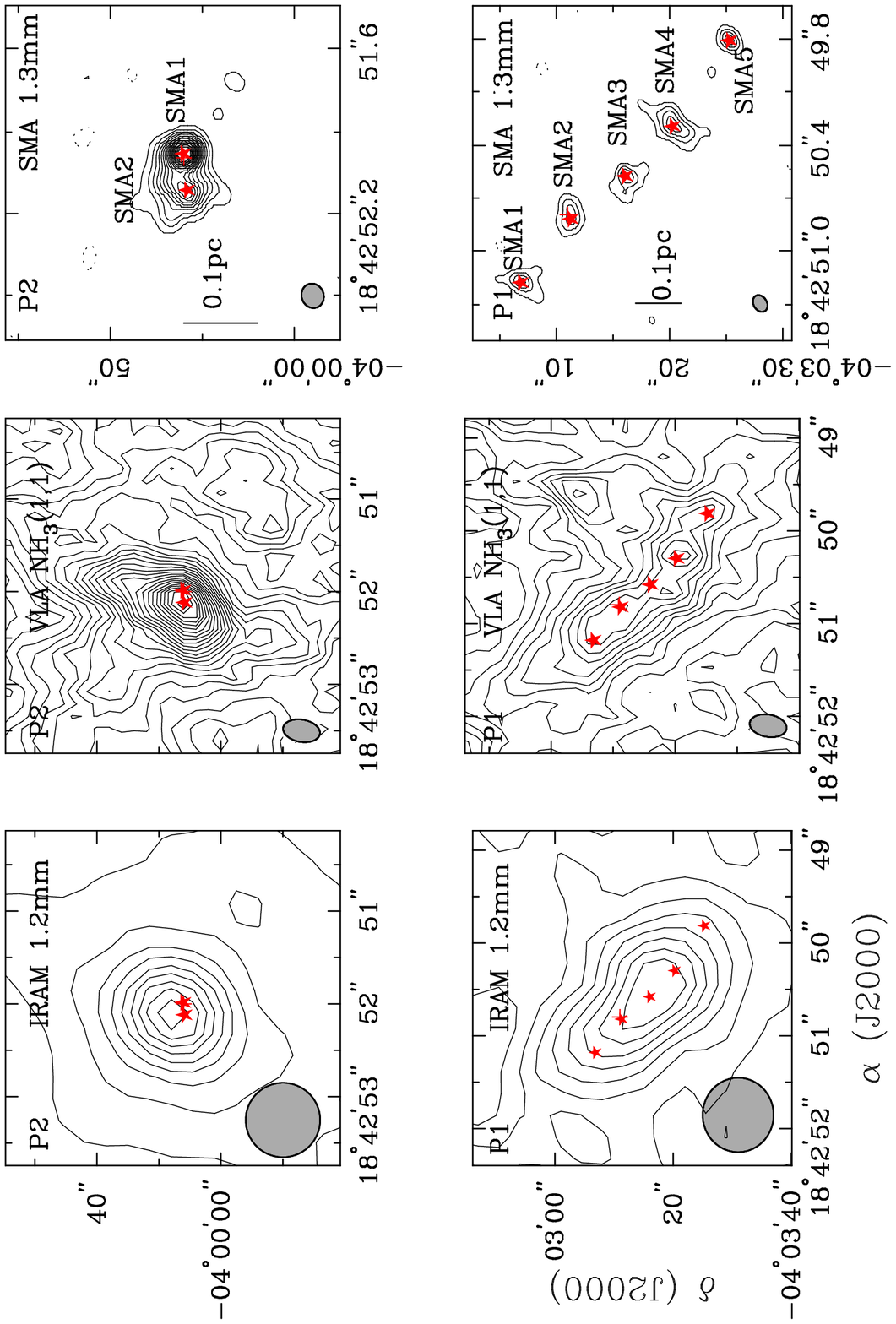}
\caption{ }
\end{figure}

\newpage

\begin{figure}[h]
\figurenum{3a}
\includegraphics{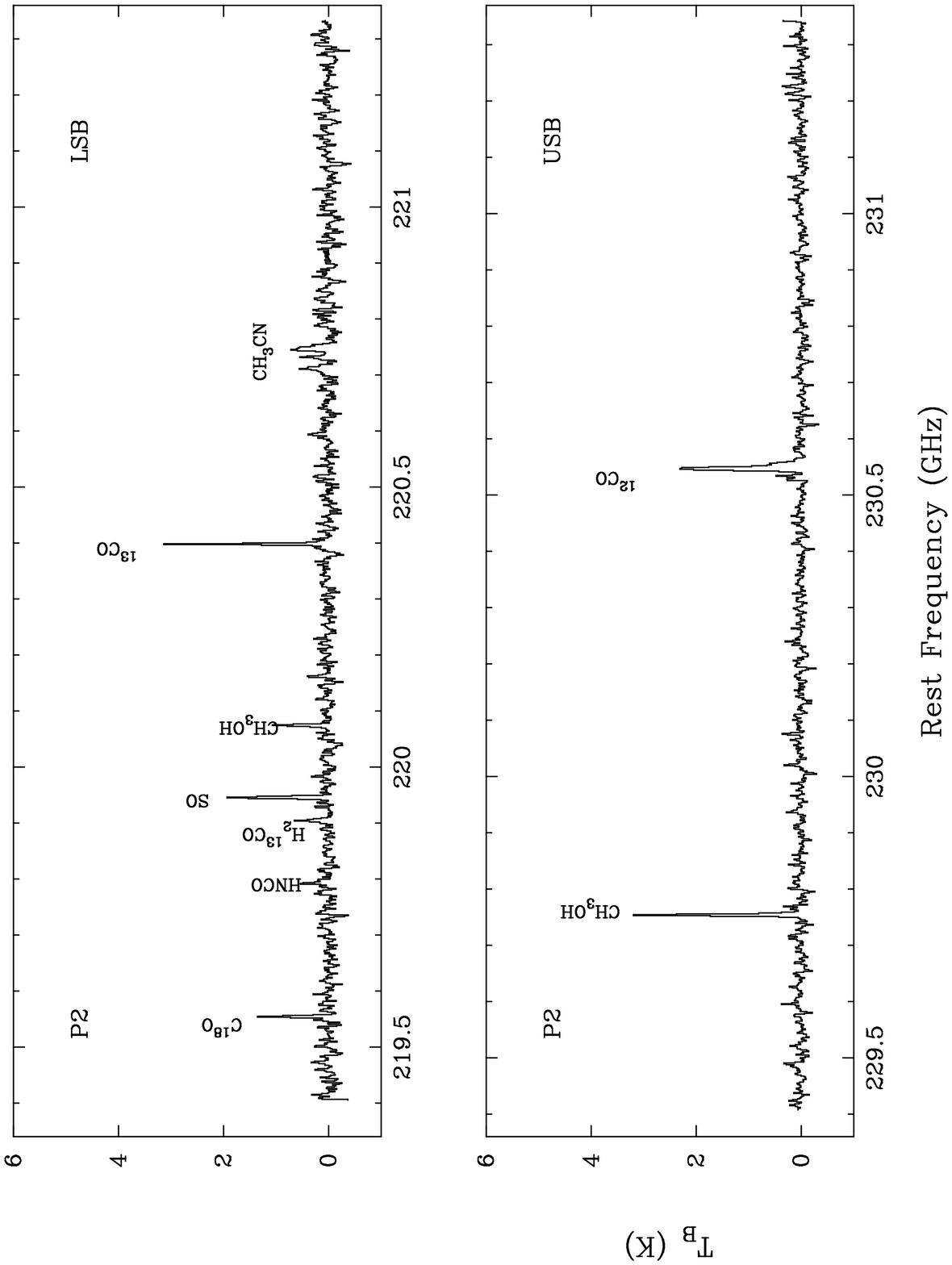}
\caption{ }
\end{figure}

\newpage

\begin{figure}[h]
\figurenum{3b}
\includegraphics{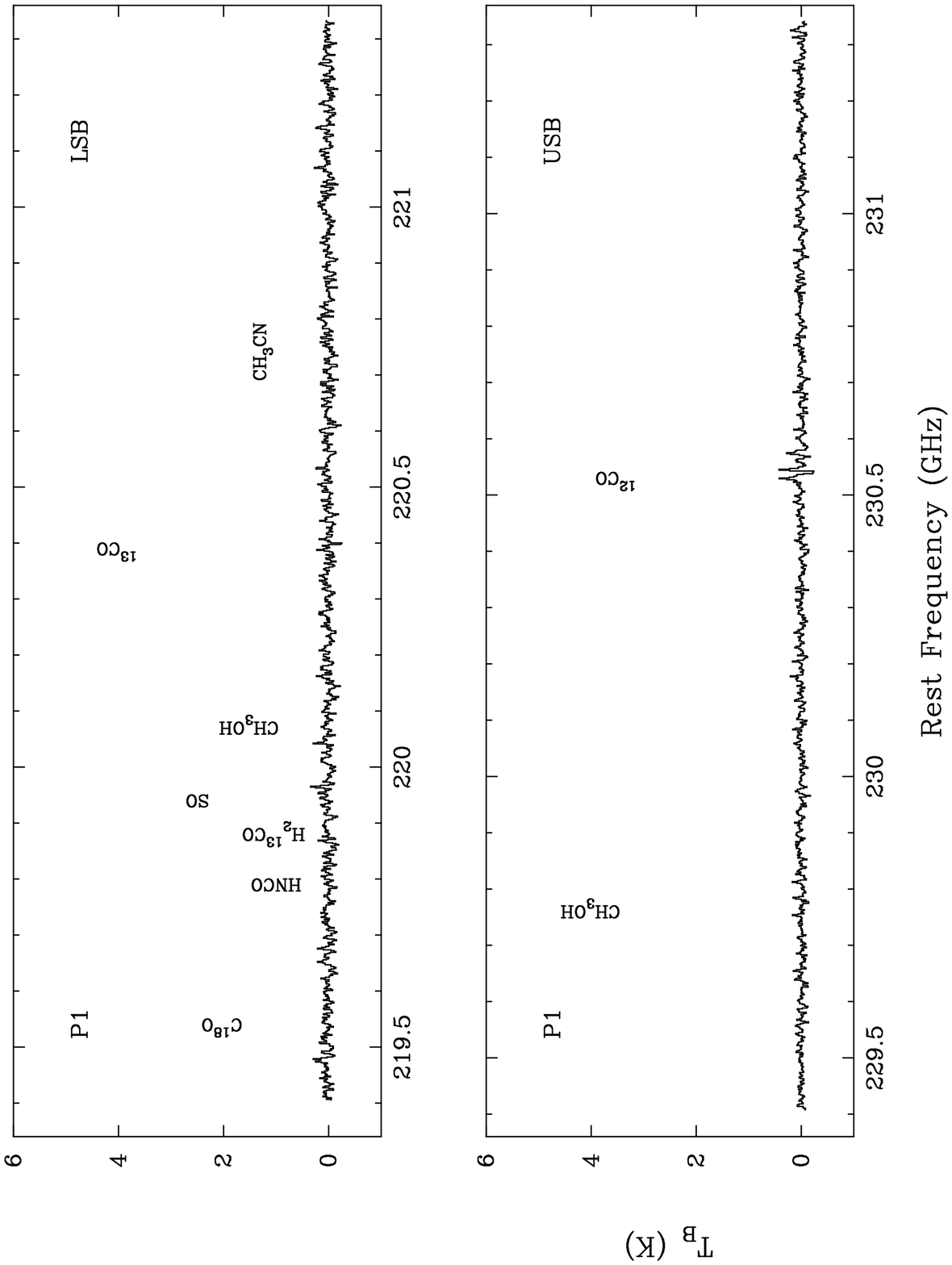}
\caption{ }
\end{figure}

\newpage

\begin{figure}[h]
\figurenum{4}
\includegraphics{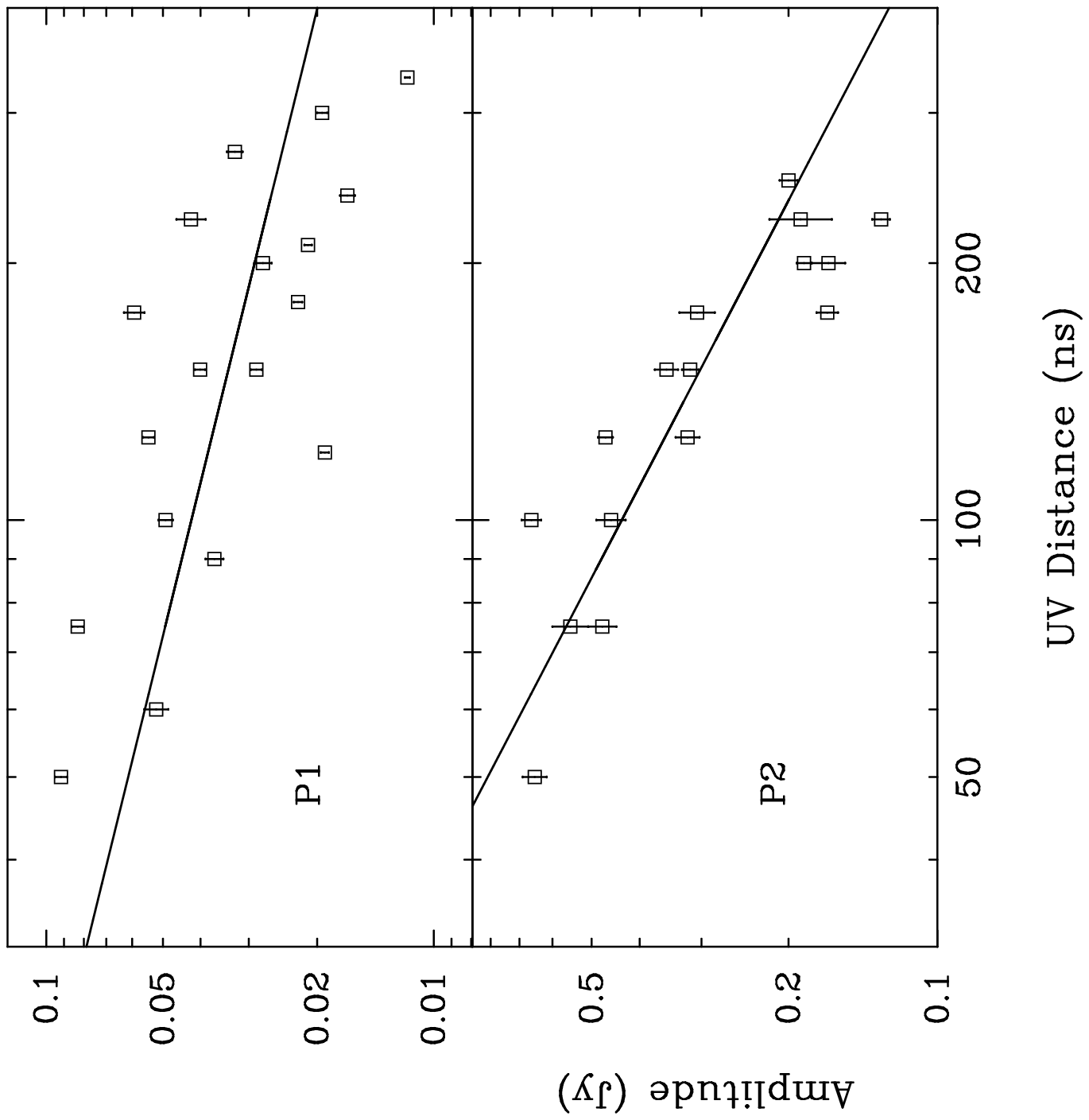}
\caption{ }
\end{figure}

\newpage


\end{document}